# *In-situ* Investigation of the Early Stage of TiO$_2$ epitaxy on (001) SrTiO$_3$

M. Radovic, [1,2] M. Salluzzo,[2] Z. Ristic,[2,3] R. Di Capua,[2,4] N. Lampis,[2,5]
R. Vaglio,[2,6] and F. Miletto Granozio[2]

[1] *LSNS - EPFL, PH A2 354 (Bâtiment PH) Station 3 CH-1015 Lausanne, Switzerland*
[2] *CNR-SPIN, Complesso Universitario Monte S. Angelo, Via Cintia I-80126 Napoli, Italy*
[3] *Département de Physique de la Matière Condensée, University of Geneva, 24 Quai Ernest-Ansermet, CH-1211 Geneva 4, Switzerland*
[4] *Dipartimento S.pe.S., Università degli Studi del Molise - Via De Sanctis, I-86100 Campobasso, Italy,*
[5] *Laborvetro srl, via Calamattia 6, I-09134 Cagliari, Italy*
[6] *Dipartimento di Fisica, Università "Federico II" di Napoli, Piazzale Tecchio, I-80126, Italy*

We report on a systematic study of the growth of epitaxial TiO$_2$ films deposited by pulsed laser deposition on Ti-terminated (001) SrTiO$_3$ single crystals. By using *in-situ* reflection high energy electron diffraction, low energy electron diffraction, x-ray photoemission spectroscopy and scanning probe microscopy, we show that the stabilization of the anatase (001) phase is preceded by the growth of a pseudomorphic Sr—Ti—O intermediate layer, with a thickness between 2 and 4 nm. The data demonstrate that the formation of this phase is related to the activation of long range Sr migration from the substrate to the film. The role of interface Gibbs energy minimization, as a driving force for Sr diffusion, is discussed. Our results enrich the phase diagram of the Sr—Ti—O system under epitaxial strain opening the route to the study of the electronic and dielectric properties of the reported Sr-deficient SrTiO$_3$ phase.

Keywords: Titanium dioxide, thin film growth, oxides, PLD

## INTRODUCTION

Oxide interfaces have boosted in recent years an unprecedented interest due the novel functional properties that are not present in the single constituting layers. A widely celebrated example is provided by the discovery of a two-dimensional electron gas at LaAlO$_3$/SrTiO$_3$ interfaces [1]. The exciting novel properties of this system were explained in terms of a purely "electronic reconstruction" mechanism [2]-[3]. However, extrinsic doping provided by point defects may have an equally important role, as it might occurs in an intermixed interface. In this framework, cation intermixing at the SrTiO$_3$ (STO) surface is a central issue. Cation interdiffusion in all-perovskite heterostructures is in general modest, and recent data [4] set a single atomic plane as the upper limit to the amount of cations crossing the interface. Nevertheless, intermixing and structural roughing of the interfaces, even occurring at the level of a single atomic plane, can effectively affect the functional properties of these systems [5]. Consequently, studies on the early stage of growth of oxide thin films on STO single crystals are extremely important in a more general context.

In this paper we present a detailed study of the early stages of the growth of anatase TiO$_2$ (001) thin films deposited by pulsed laser deposition (PLD) on TiO$_2$ terminated SrTiO$_3$ (001) single crystals. The study of the growth of TiO$_2$ is motivated by the recent interest in photocatalysis [6], solar cells [7], gas sensors [8], photovoltaics [9], and more recently, in spintronics applications [10]. While the growth of both anatase and rutile TiO$_2$ films was successfully accomplished on LaAlO$_3$ [11] and untreated SrTiO$_3$ [12] single crystals, a throughout investigation of the early stage of growth of this material on well-ordered TiO$_2$ terminated (001) SrTiO$_3$ surfaces, to our knowledge, was not accomplished since now.

The ideal crystal structure of epitaxial TiO$_2$/SrTiO$_3$ system is shown in fig.1a. An STO (001) crystal can be envisaged as a vertical a stack of alternating SrO (A-type) and TiO$_2$ (B-type) planes. The TiO$_2$ planes in STO are flat and undistorted, with 0.3905 nm lattice spacing. They are in tensile strain, due to the intercalated SrO planes, a circumstance that makes STO an incipient ferroelectric [13]. The anatase phase of TiO$_2$ is, instead composed by a vertical stack of alternated TiO$_2$ (B-type) planes. The TiO$_6$ octhaedra in the anatase structure are distorted and rotated one respect the other. Consequently each plane is strongly buckled (with O-Ti-O bond angle $\alpha \approx 156°$).

Since the misfit between the in plane lattice of (001) TiO$_2$ anatase (a=b =0.3785 nm) and SrTiO$_3$ amounts to 3% (Table 1), epitaxial films are expected to be under a tensile strain. Due to this large lattice mismatch, the stability of a coherent anatase TiO$_2$(001)/STO(001) interface, implying straightening and stretching of the Ti-O bonds, is expected to have a high cost in terms of elastic energy. The stability of such distorted structure, composed of a stack of many straightened/stretched TiO$_2$ planes [14], is obviously questionable. Several strain-relief mechanisms can take place, including the relaxation of the TiO$_2$ layers by island growth, induction of oxygen non-stoichiometry, cation interdiffusion, or others. In particular, cation interdiffusion over many layers could play a crucial role in the case of the anatase/perovskite structure, which is opposite to what is expected for all-perovskite heterostructures where a long-range diffusion of large cations (as A-site La or Sr) through interstitial sites is inhibited. Additionally, we notice that A-site cation intermixing between two stoichiometric perovskites requires the double exchange of A-type cations, which can be considered to some extent as a "second order" process. For this reason, the properties of perovskite oxide interfaces are in general affected by the more energetically favorable formation of oxygen vacancies [15]. On the other hand, the interlayer diffusion from the A-type sub-lattice of STO into the anatase lattice might be favored both by entropy and strain minimization, and should be envisaged, as a "first order" process.

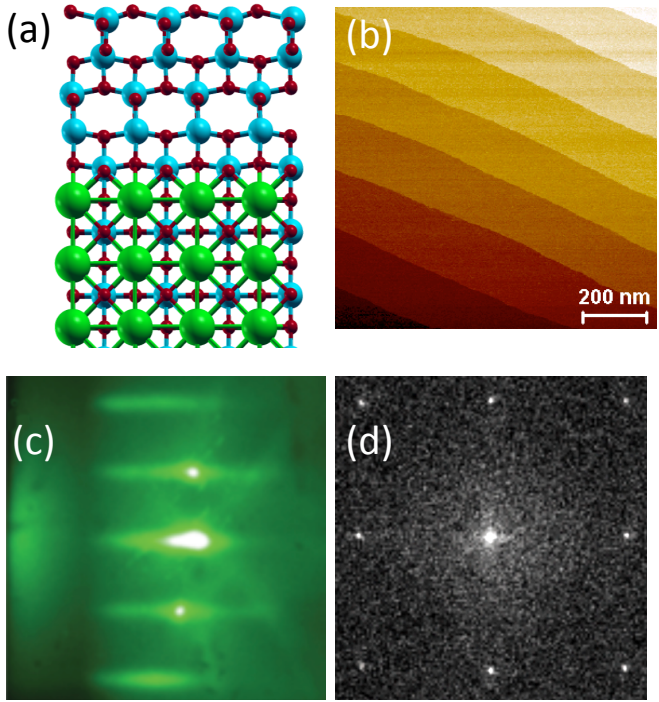

Fig. 1: Ideal structure of a coherent growth of strained anatase $TiO_2$ (001) film on $TiO_2$ terminated $SrTiO_3$ single crystal (Sr ions are green spheres, Ti are cyan and oxygens ions are in red). (b) *In-situ* non-contact AFM surface topography of our Ti terminated STO substrates. (c) RHEED and (d) LEED patterns of single terminated samples.

On the base of the former qualitative considerations, it can be argued that interdiffusion of Sr ions into $TiO_2$ is a likely process. Therefore, it might either induce the formation of an inhomogeneous, phase separated, film or favor the epitaxial growth of a non-stoichiometric pseudomorphic Sr deficient $Sr_xTiO_{2+y}$ phase (x, y <1). The latter possibility is very interesting, since the phase diagram of the Sr–Ti–O system does not contain any thermodynamic phase with a [Sr]/[Ti] ratio < 1 [16]. On the other hand, the electronic properties of a $SrTiO_3$ phase in the Sr-deficient unexplored region of the phase diagram can be interesting for possible applications.

In the following, we report on a comprehensive all-*in-situ* study of the earliest stages of growth of $TiO_2$ on STO single crystals. We show that an intermediate, Sr-deficient, strain-stabilized, perovskite-like phase with a thickness extending up to 2-4 nm, can develop at the $TiO_2$/STO interface in controlled growth conditions. The formation of such phase corresponds to the self-organization of the interstitial Sr, diffused into the epitaxially strained $TiO_2$, in a partially filled A-site sublattice. Such behavior is not found when $TiO_2$ films are deposited on other substrates, like $LaAlO_3$ (LAO) and $SrLaAlO_4$ (SLAO), where anatase grows respectively in null or slightly compressive (-0.8%) strain [Table 1]. The realization of this phase is confirmed by detailed ex-situ microscopic and compositional analysis, which will be reported in a forthcoming paper [17].

**EXPERIMENTAL RESULTS**

The multi-chamber ultra high vacuum (UHV) apparatus employed in this work was specifically designed for pulsed laser deposition of epitaxial oxide thin films and multilayers and for the analysis of the structural and electronic properties of the fresh uncontaminated surfaces of as-grown samples. It includes, (i) an UHV PLD setup equipped with a KrF excimer laser and a 30 KeV high pressure RHEED; (ii) an analytical chamber dedicated to photoemission (XPS, UPS), low energy electron diffraction (spot profile analysis LEED) and scanning probe microscopy (SPM) and spectroscopy analyses. $TiO_2$ ultra-thin films were grown on (001) $SrTiO_3$, $SrLaAlO_4$ (SLAO) and $LaAlO_3$ ($LaAlO_3$) single crystals and analyzed in situ, resorting to the above mentioned complementary surface science techniques. In the case of STO single crystals, we employed both untreated (equally SrO and $TiO_2$ terminated) and single $TiO_2$ terminated substrates. The latter were prepared following a well-established etching procedure [18], and subsequently annealed *in-situ* in the PLD chamber in 0.1 mbar of $O_2$ at 950°C for 2h right before deposition. The in-situ annealing procedure was adopted to reduce the risk of any possible surface contamination. RHEED, LEED, AFM and surface diffraction analyses [19] of these single crystals demonstrate their high quality and a single termination [fig.1b-d].

$TiO_2$ films were grown by PLD in flowing 0.1 mbar molecular oxygen using a stoichiometric ceramic $TiO_2$ target, typical laser fluencies of 2.5 J cm$^2$, and substrate temperatures of 800°C. Some test samples were also deposited at different temperatures in the 650°C-800°C range, yielding substantially similar results. One sample intentionally was grown at room temperature for reference. Grazing incidence x-ray diffraction was used to calibrate the deposition rate of $TiO_2$ in our typical deposition conditions, which corresponds to about 55 pulses/nm or, equivalently, about 14 pulses/$TiO_2$ monolayer. Further data are also reported in [20]. Films deposited on LAO and SLAO exhibit the typical (4x1)x(1x4) surface reconstruction of anatase $TiO_2$ (001) films [21]. These samples exhibit a high degree of crystallographic perfection and relatively smooth surfaces (rms roughness =

Table 1: calculated mismatch between anatase (001) TiO$_2$ and (001) STO, SLAO and LAO substrates; fitting values according to eq. 1 of the parameter t$_0$ for films deposited on different substrates.

| Type of Substrate | STO etched (001) | STO non-etched (001) | SLAO (001) | LAO (001) |
|---|---|---|---|---|
| Mismatch % | -3% | -3% | +1% | -0.3% |
| t$_0$ (nm) | 0.8 | 0.22 | <0.2 | <0.2 |

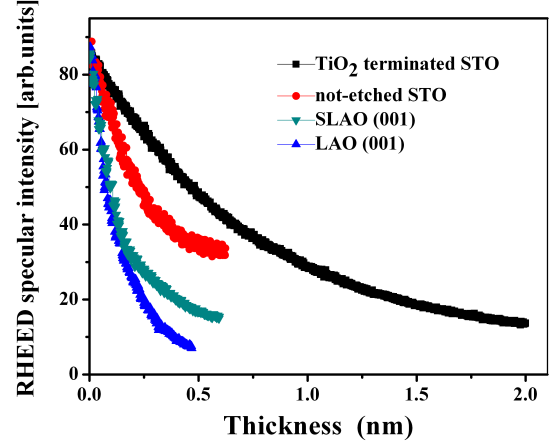

Fig. 2 Linear-log plot of the exponential decay of the specular RHEED spot intensity vs. nominal film thickness for TiO$_2$ films deposited on TiO$_2$ terminated STO (black squares), on non-etched, equally SrO and TiO$_2$ terminated STO (red circles), SLAO (green down triangles) and LAO (up blue triangles) substrates.

2nm on 1x1 μm$^2$ area), as confirmed by RHEED, LEED and STM measurements [20]. However, the TiO$_2$ films on STO (001) single crystals are characterized by a strikingly different behavior. In particular the growth-mode of TiO$_2$ is strongly affected by the crystal quality and strongly depends on the atomic termination of the substrate surface.

We will now draw our attention on data obtained either on high quality surfaces of nominally TiO$_2$ terminated STO or on homo-epitaxial STO thin films. The first striking feature that was observed for the growth on Ti-terminated STO, was a steady RHEED pattern that remained unchanged during the first nominal 2-4 nm of TiO$_2$ deposition, i.e 8-16 TiO$_2$ atomic layers (one TiO$_2$ atomic layer =0.24 nm). This is in contrast with TiO$_2$ growth on LAO and SLAO, where the RHEED pattern quickly changed on time scales corresponding to the deposition of 1-2 monolayers (fig. 2). The unique real-time indication of a thin film growth was a slow exponential decrease of the specular RHEED intensity (fig. 2). The data can be very well fitted using a single exponential,

$$I(t) = I_0 + ke^{-t/t_0} \qquad (1)$$

Samples grown on LAO and SLAO are characterized by typical values of the parameter t$_0$ of the order 0.15 nm (7-8 laser pulses) [22]. This value increases to 0.24 nm (13-14 pulses) for samples deposited on untreated STO, but strikingly go up 0.8-0.9 nm (44-50 pulses) when TiO$_2$ is deposited on TiO$_2$ terminated STO or on STO films (table 1).

The "STO-like" RHEED pattern remained unchanged until a thickness of few nanometers is deposited. The phenomenon was reproducibly found on several samples. We routinely employed the very regular RHEED oscillations of homoepitaxial STO films grown on Ti-terminated STO as an independent calibration of our laser pulse energy on the target. Such procedure guarantees that comparable growth conditions hold for all samples. The early stage of the growth of a TiO$_2$ on such as grown STO layers showed the same phenomenology. These real time analyses suggest that the early stage of growth of TiO$_2$ on a TiO$_2$ terminated SrTiO$_3$ takes place by forming a pseudomorphic epitaxial layer, growing in step-flow mode.

We performed systematic in-situ studies of these pseudomorphic films by using RHEED, LEED, SPM and XPS techniques. RHEED (fig. 3a) and LEED (fig. 3b) demonstrate that these layers are characterized by an in-plane structure indistinguishable from the one of the STO single crystal substrate. In particular we can notice the very sharp LEED diffraction spots, which suggest an in plane structural coherence that is comparable to that one of the substrate. Moreover, these data allowed to precisely determine the in plane lattice parameters, which within 1% were consistent with the square STO 2D unit cell of 0.39x0.39 nm$^2$. STM data show that this pseudomorphic phase is composed by well ordered flat terraces thus confirming a step-flow growth mode with TiO$_2$ ad-atoms incorporated at the STO step edges [fig. 3c]. The height of the terraces, being c=0.37±0.01 nm, very surprisingly are much closer to the (001) STO unit cell than to any possible characteristic distance among consecutive equivalent planes in the rutile or anatase phases of TiO$_2$.

Therefore both LEED and STM data suggest that the ultra-thin pseudomorphic TiO$_2$ films resemble in many respects the structural characteristics of an "STO-like" pseudo-cubic phase. The surface morphology, although qualitatively similar to the STO substrate on which these films were grown, shows faceting along {100} and {010} directions that is not typical for STO (see for example the growth features of homoepitaxial STO shown in [23]), suggesting that (100) surfaces have the lowest surface energy. Driven by the analogy with Bravais cubic lattices, we speculate that this might be attributed to a partial occupancy of the central A site of the cubic cell. Indeed, (110) surfaces have the lowest energy in BCC structures, while (100) surfaces are the lowest energy ones in simple cubic systems.

Having established that a thin epitaxial pseudomorphic film is formed during the early stage of growth, we tried to obtain information about the stoichiometry of this phase. As shown in fig.2, the intensity of the RHEED specular spot follows an exponential decay. Taking into account that the surface morphology and the structural perfection appear similar to those one of the STO substrate, this result might be possibly related to a stoichiometry deviation from pure, $TiO_2$ terminated, STO. In particular, the RHEED diffracted intensity could decrease as function of the thickness if the composition of the element with the highest atomic form factor decreases with the time, i.e. Sr in STO. A demonstration of a change of Sr-stoichiometry comes from the XPS analyses, allowing an estimation of the Sr/Ti ratio by comparing the integrated intensities of the Sr $3d_{5/2}$-$3d_{3/2}$ and Ti $2p_{3/2}$-$2p_{1/2}$ photoemission lines of different samples. We acquired spectra in two geometrical configurations, normal emission (90° to the substrate plane) and grazing emission, (25° to the substrate plane). Since the estimated inelastic mean free path $\lambda$ is $\approx 1.4$ nm for Sr 3d photoelectrons (1119 eV kinetic energy for our Mg K$\alpha$ source) the grazing configuration probes a depth of the order of the film thickness (1-2 nm). Fig. 4a shows a comparison between the Sr $3d_{5/2}$-$3d_{3/2}$ photoemission peaks of untreated STO single crystal (having equal amount of Sr and Ti at the surface and in the bulk) and of a pseudomorphic 3 nm $TiO_2$ film deposited on a $TiO_2$ terminated STO (pseudomorphic phase). Data on untreated STO, 3nm anatase film (deposited on untreated STO) and 3nm amorphous $TiO_2$ (grown at room temperature) are also shown for comparison. We notice that the "STO-like" pseudomorphic sample is characterized by a relatively high Sr content (quantitative evaluation indicate a stoichiometric ratio Sr/Ti $\approx 0.7$), which is slightly lower than the Sr content of $SrTiO_3$. On the other hand, as expected, the anatase and amorphous film show a much lower Sr signal, which is due only to the contribution of the $SrTiO_3$ single crystal below. Such contribution, indeed, disappears for a 10 nm thick film, which is predominately composed by a $TiO_2$ anatase phase.

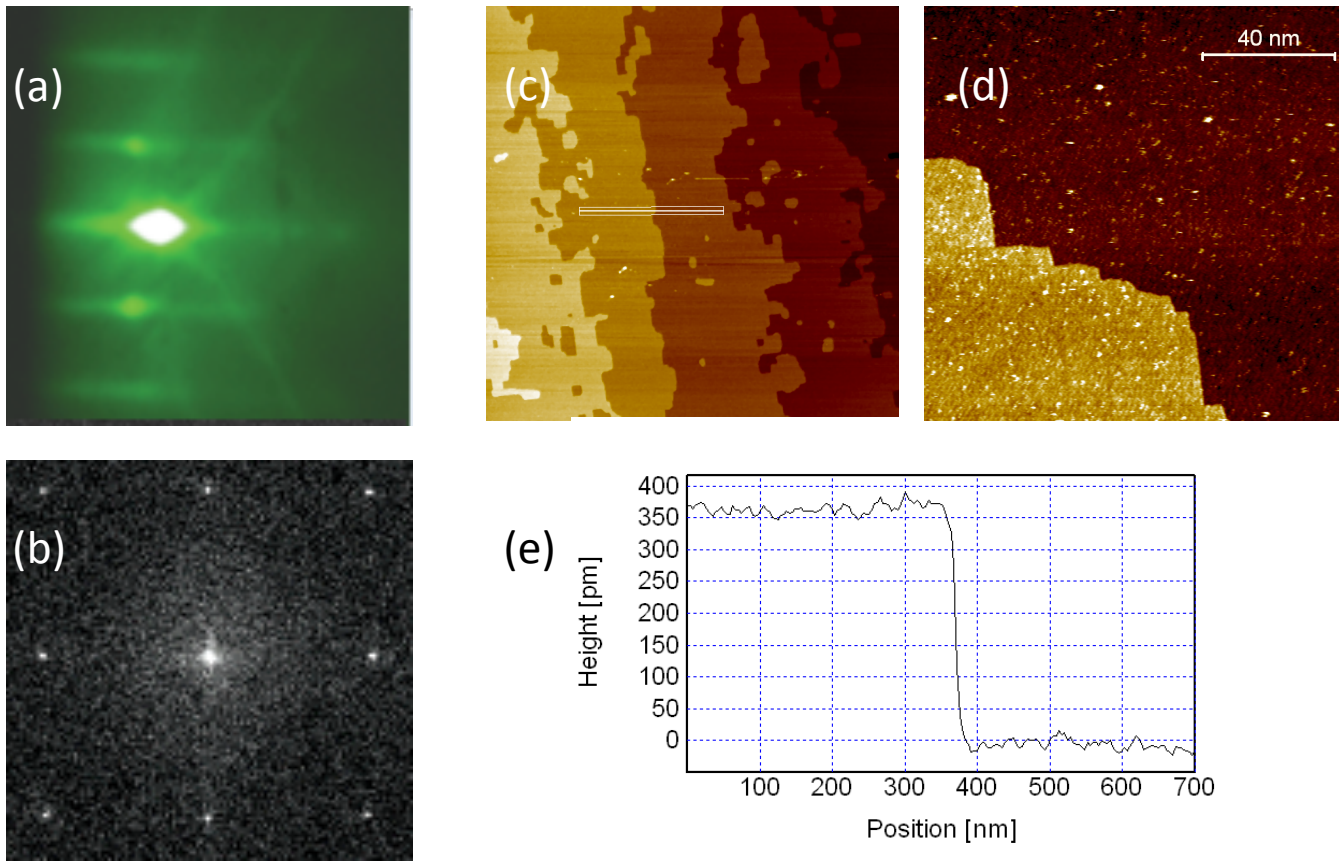

Fig. 3: Typical (a) RHEED (Specify energy) and (b) LEED (E=120 eV) patterns (c, d) STM surface topography of our 2nm $TiO_2$ thin film grown on $TiO_2$ terminated STO substrates. RHEED and LEED patterns are unchanged with respect to the previous Ti terminates STO surface. The terrace structure also persists, showing faceted edges along the in plane principal directions, absent in the case of STO surfaces (d). The step profile (e) shows a vertical lattice parameter which is close to the STO c-axis.

Thus the combination of RHEED, LEED, STM and XPS in-situ techniques allows establishing that a $Sr_xTiO_{2+y}$ pseudomorphic STO-like phase is realized in the early stage of PLD growth of $TiO_2$ films on $TiO_2$ terminated STO substrates. Additionally, a transition to the anatase phase occurs above 10 nm.

The nucleation of the $TiO_2$ anatase phase, is easily identified by RHEED in real time with the appearance of a typical 3D pattern, which shows the transition from the step-flow growth mode characteristic of the pseudomorphic, "STO-like", phase, to a 3D island growth mode [fig.5]. In order to analyze more clearly the transition between pseudomorphic to anatase phases, we studied the three different stages of the deposition process (pseudomorphic, intermediate, anatase) by moving back and forth a single sample from the PLD chamber to the analytical chamber. The LEED pattern of the intermediate phase (fig. 6a) is composed by sharp diffractions spots associated to a square lattice similar to the "STO-like" phase, but with lattice parameters about 2-3% smaller, and a new "cross-like" structure suggestive of a six-fold reconstruction. AFM measurement (fig.6b-6c) clarifies that the surface is composed by small 3D-islands on top of flat terraces. Thus, in this intermediate phase, the sample is composed by a major "STO-like" phase and a minority of 3D islands, which on the other hand give rise to the typical 3D RHEED pattern of fig.5b, due to the grazing incidence conditions. The appearance of 3D islands determines the end of the "STO-like" deposition phase, and the subsequent incorporation of $TiO_2$ ad-atoms to the islands, is interpreted on the base of our data as the nucleation of a relaxed anatase. The Sr migration into the film is expected to be, in this case, mostly suppressed with respect to the perovskite-like phase, due to the absence of available A-type sites in the $TiO_2$ structure. Upon further $TiO_2$ deposition, these islands coalesce and recover a predominant 2D surface. The anatase film is easily recognized from the appearance of the typical 4x1 surface reconstruction, as shown in fig.7, where extensive RHEED and LEED and STM data allow to demonstrate that the reconstruction is related to the the presence of a two-domain row-like structure on the anatase surface (fig. 7c). The presence of domains can be attributed to the coalescence of islands with different row alignments. A single domain area is shown in fig. 7d.

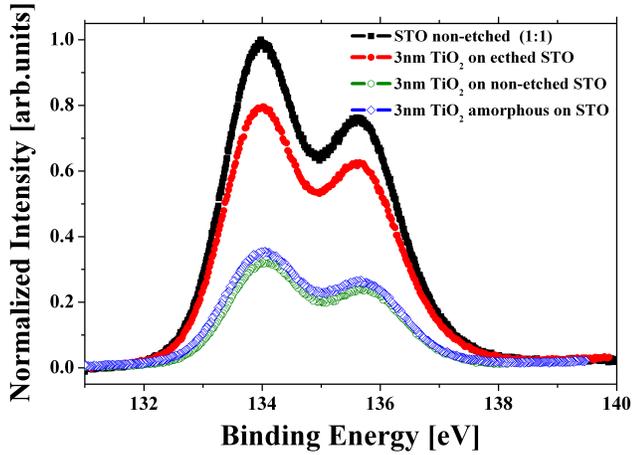

Fig. 4 : (a) $3d_{5/2}$-$3d_{3/2}$ Sr peaks of a untreated (equally SrO and $TiO_2$ terminated) STO single crystal (black line), and 3 nm (nominal) $TiO_2$ samples deposited on it (blue line) and on a $TiO_2$ terminated STO (red line). The intensities of all the 3d5/2-3d3/2 Sr peaks were normalized to the $2p_{3/2}$-$2p_{1/2}$ Ti emission. All data have been rescaled by setting the maximum of the STO Sr emission intensity to 1.

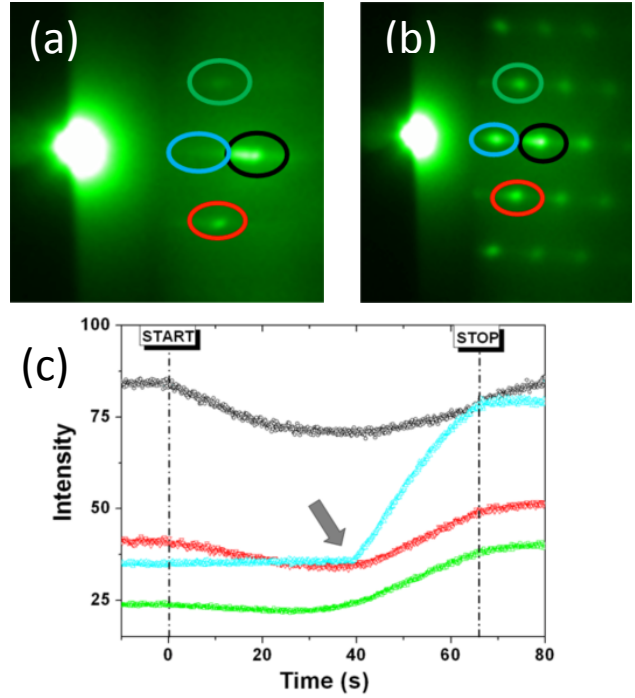

Fig. 5 RHEED pattern of a $TiO_2$ films during (a) the pseudomorphic growth of the STO-like phase, and (b) during the intermediate phase where 3D anatase islands appear. In (c) the RHEED intensity of different diffraction spots as function of time is shown (the colors corresponds to the circles in (a) and (b)). The arrow indicate the starting of nucleation of 3D islands.

## DISCUSSION

It is widely recognized that under reducing conditions, STO undergoes several complex chemical reactions involving defect formation and diffusion. In brief, two main processes take place. The first one is the creation of oxygen vacancies [24][25][26]. The second, slower reaction involves the Sr cations. It was at first proposed that this is achieved by the establishment of rocksalt intergrowth layers (the Ruddlesden-Popper phases $SrO_x(SrTiO_3)_n$) [27]. However, no experimental indication of such intergrowth in $SrTiO_3$ has ever been found. On the contrary, overwhelming evidence indicates an alternative defect mechanism, where strontium vacancies and the SrO-rich second phase can be exclusively created at the surface of single crystals due to cation diffusion [28, 29, 30].

Once the Sr migration at the STO surface is accepted, it is possible to propose a model that accounts for the whole body of our results regarding the $TiO_2$ growth. The reaction of creations of Sr vacancies and SrO under the growth conditions may then be written as:

$$\frac{1}{2}O_2(g) + 2e' + Sr_{Sr}^{\times} \xrightarrow{ox} V_{Sr}^{''} + SrO \qquad (2)$$

where $Sr_{Sr}^{\times}$ and $V_{Sr}^{''}$ represent metal ion and metal vacancy. At the deposition conditions, Sr vacancies are formed close

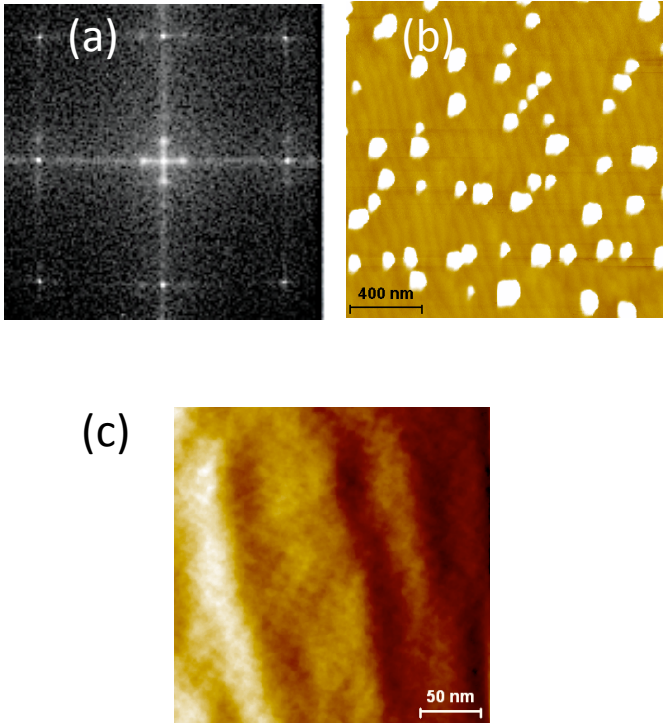
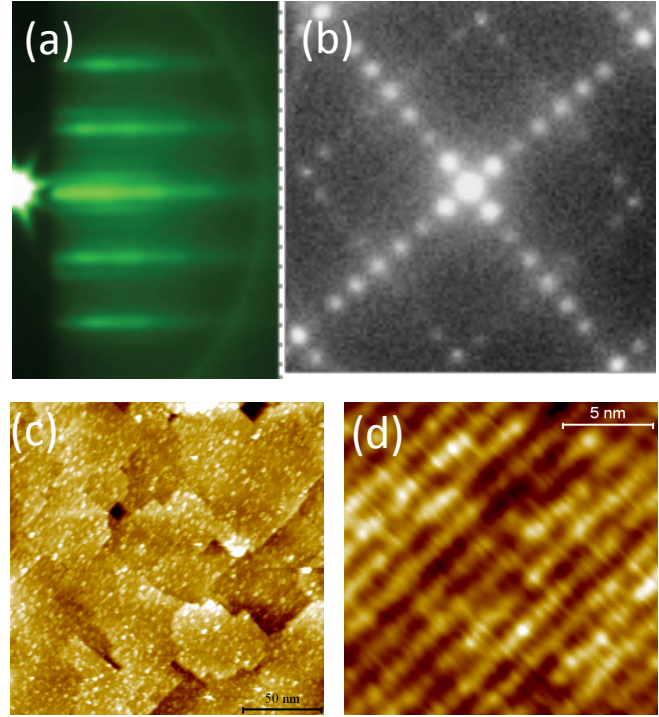

Fig. 6 (a) LEED pattern at the transition between phase B and phase C. The in plane lattice parameter, as deduced by the dimension of the BZ, shows a ≈ 2-3% reduction in real space, compatible with a partial relaxation towards the $TiO_2$ in plane lattice parameter values. Furthermore, a cross-like structure compatible with a six-fold reconstructions is visible around the diffractions spots. (b) AFM image showing the nucleation of islands on a background surface were the presence of steps is still visible. Islands have lateral dimensions of the order of 100nm and an average height of 3 -4 nm.(c) A zoom on a region not containing the 3D islands, showing the step structure and the presence of a sixfold superstructure.

Fig. 7: (a) RHEED and b) LEED data on as grown 10 nm thick $TiO_2$ film deposited on $SrTiO_3$ surface. (a) The RHEED pattern is indicative of the coalescence of the islands into a uniform anatase film and shows an evident 4x1 reconstruction. (b) LEED pattern showing a square in-plane lattice and the presence of two perpendicular 4x1 reconstructed domains. The spot width demonstrates an inferior crystal quality with respect to samples grown on LAO and SLAO (not shown). (c) STM images ($I_t$=1 nA Vbias =1.0 V, 200x200 $nm^2$) showing a continuous anatase film formed by the coalescence of faceted islands. A single domain area of the reconstructed surface, showing a row like structure with a periodicity of four u.c. is reported in (d). high resolution STM images of these anatase films.

to the STO surface and Sr ad-atoms appear. We believe that the growth front of the $TiO_2$ film incorporates them (Sr ad-atoms) in a continuous efficient way during the growth process, according to the following scenario: when Sr diffuses in the lattice formed by a sequence of $TiO_2$ layers, the film stoichiometry is transformed into $Sr_xTiO_{2+y}$. While the Sr-rich side of the Sr—Ti—O phase digram (Sr/Ti > 1) hosts a number of Ruddlesen Popper phases, in bulk systems Sr doped $TiO_2$ phase separates into the anatase phase, possessing a poor Sr solubility, and $SrTiO_3$ [31]. On the other hand, the formation of a defective perovskite-like structure is possible in epitaxial films, forming a $SrTi_2O_5$ composition [32].

We believe that strain minimization is the driving force that sustains the defective pseudomorphic perovskite beyond the limits assigned by a bulk phase diagram. Consequently our scenario is the following: during the pseudomorphic growth, the $TiO_2$ ad-atoms impinging on the terraces and condensing at the edges during a step-flow growth, feel a driving force to crystallize in such a form that allows a three-dimensional lattice matching, along the two in-plane directions and along the vertical one (step height). Such a matching is accomplished by resorting to the available Sr atoms and the formation of the defective perovskite. Our data suggest that, through this mechanism, a relatively small amount of intercalated Sr atoms between the two terminating $TiO_2$ layers, together with strain minimization at the step edge, can force the growing $TiO_2$ layer to align with the underlying one, straightening the Ti–O–Ti bonds and forming a cubic structure. Interstitial Sr is therefore immediately transformed in the A-site cation of a defective perovskite, having the lattice parameters matched to STO with small or null elastic energy. The formation of a new perovskite layer allows in turn the formation of a fresh step edge, where segregated Sr atoms and Ti ad-atoms can react. The essential role of strain in sustaining this mechanism is proven by the fact that films relaxing to the anatase phase in the very first stages of growth have a much lower Sr content (fig. 5), and that long high temperature post-deposition annealing were shown by XPS to have a sizable but modest effect in increasing the Sr content at the surface. Although the STO substrate can be considered as an infinite Sr reservoir, the limited mobility will cause a decrease of the Sr content with the thickness. The process will stop abruptly with the nucleation of unstrained anatase islands, which will cover the surface eventually giving rise to an uniform $TiO_2$ anatase film.

**CONCLUSIONS**

We showed that, differently from the case of perovskite/perovskite interfaces, a long range cation interdiffusion process can be activated at the anatase (001) $TiO_2$/(001)$SrTiO_3$ interface. By adopting an all-*in-situ* approach for monitoring the surface of the growing samples by several complementary techniques, we show that the transition from the initial Ti terminated STO surface to the final uniform $TiO_2$ films takes place through two intermediate phases: the growth of a strain stabilized, lattice-matched defective $Sr_xTiO_{2+y}$ perovskite, due to the interdiffusion of Sr from the substrate, and the nucleation on the previous phase of three dimensional relaxed anatase islands.

Our results, beside pointing out to the need of resorting to alternative substrate materials for the growth of high quality anatase $TiO_2$ films, enrich the phase diagram of the Sr—Ti—O system under epitaxial strain opening the route to the study of the electronic and dielectric properties of the reported Sr-deficient STO phase.

**ACKNOWLEDGMENTS**

The authors wish to thank U. Scotti di Uccio and G. Cantele for discussions and for the manuscript preparation, and A. Falqui for the cooperation in this research. N. L. acknowledges financial support from the Autonomous Region of Sardinia through a research grant under the program "Promoting scientific research and innovation technology in Sardinia".

The research leading to these results has received funding from the European Union Seventh Framework Programme (FP7/2007-2013) under grant agreement n° N. 264098 - MAMA.